\documentclass{emulateapj}
\usepackage{color}
\usepackage{rotating}
\usepackage{gensymb}

\shorttitle{Sextans: core, cusp or MOND?}
\shortauthors{Lora et al.}
\begin{document}
\title{Sextans' cold substructures as a dynamical judge: Core, Cusp or MOND?}
\author{V. Lora\altaffilmark{1}, E. K. Grebel\altaffilmark{1}, F. J. S\'anchez-Salcedo\altaffilmark{2}
and A. Just\altaffilmark{1}}
\altaffiltext{1}{Astronomisches Rechen-Institut, Zentrum f\"{u}r Astronomie der Universit\"{a}t Heidelberg, \\
             M\"{o}nchhofstr. 12-14, 69120 Heidelberg, Germany}
\altaffiltext{2}{Instituto de Astronom\'{\i}a,
              Universidad Nacional Aut\'onoma de M\'{e}xico,
              AP 70-264, 04510 D.F., M\'{e}xico}
\email{vlora@ari.uni-heidelberg.de}


\begin{abstract}

The cold dark matter model predicts cuspy dark matter halos. However, it has been found 
that, in some low-mass galaxies, cored dark halos provide a better description of their
internal dynamics.
Here we give constraints on the dark halo profile in the Sextans dwarf spheroidal galaxy
by studying the longevity of 
two cold kinematic substructures detected in this galaxy.
We perform $N$-body simulations of a stellar clump in the Sextans dwarf galaxy, including a
live dark matter halo and the main stellar component.
We find that, if the dark halo is cuspy, stellar clumps orbiting with semi-major
axis $\approx 400$ pc are disrupted in $\sim5$~Gyr, even if the clump is initially as compact
stellar cluster with a radius of $r_c=5$~pc. Stellar clusters in an initial orbit with
semi-major axis $\leq 250$ pc may survive to dissolution but their
orbits decay towards the center by dynamical friction.
In contrast, the stellar clumps can persist for a Hubble time 
within a cored dark matter halo, even if the initial clump's radius is as extended as
$r_c=80$~pc.
We also study the evolution of the clump in the MONDian context. In this scenario, we find that
even an extended stellar clump with radius
$r_c=80$~pc survives for a Hubble time, but an unrealistic
value for the stellar mass-to-light ratio of $9.2$ is needed.

\end{abstract}


\section{Introduction}

The $\Lambda$ cold dark matter ($\Lambda$CDM) model has 
proved to be successful in reproducing structure formation at large scales, but it 
faces some difficulties at galactic scales. 
For example, cosmological $N$-body simulations predict halos with a
central cusp \citep{navarro97,moore99,jing00}, whereas observations of the rotation curves
of dwarf and low surface brightness galaxies indicate that a cored dark halo
is preferred \citep{bosch00,blok02,kuzio08,blok08,donato09}. 

Dwarf spheroidal (dSph) galaxies are the natural targets to study the properties of dark matter 
(DM) halos at very small masses. The analysis of the cusp-core controversy in dSph galaxies
has motivated much work \citep{kleyna03,goerdt06,sanchez06,gilmore07,battaglia08,walker11,jardel12,salucci12,
agnello12,amorisco13,breddels13}. Dynamically cold stellar substructures as those observed in
some dSph galaxies are sensitive probes of the gravitational potential.
Kleyna et al. (2003) presented evidence that the stellar substructure in the Ursa Minor (UMi) dwarf
spheroidal is incompatible with a cusped DM halo. They argued that 
the second peak located on the north-eastern side of the major axis of UMi
is a disrupted stellar cluster that has survived in phase-space 
because the underlying gravitational potential is close to harmonic \citep{kleyna03,read06,sanchez-lora07}. 
This implies that the dark halo in UMi should have a cored mass density profile, 
 instead of a cuspy mass profile as that predicted by the $\Lambda$CDM model, 
and that this core should be large \citep{kleyna03,lora09}. \cite{lora12} showed that the clump 
in UMi would be short lived if the dark halo were strongly substructured, but that it can survive 
for a Hubble time in a smooth halo with a large core.

There is also evidence of the existence of stellar kinematically cold substructures 
in the Sextans dSph galaxy. \cite{kleyna04} found a drop in the stellar velocity dispersion
at their innermost data bin, which was interpreted as
a dissolving cluster at the Sextans center. Later on, \cite{walker06} 
detected a region near Sextans’ core radius that appeared kinematically colder than the 
overall stellar population of Sextans, but they did not detect any signs of a  kinematically
distinct population at the center of Sextans.
Recently, \cite{battaglia11} reported the detection of a cold substructure of
very metal-poor stars close to the Sextans center. It remains unclear if this substructure
is the same as that previously found in \cite{kleyna04}.

Our main aim is to see if the longevity of the cold substructure found in Sextans
can shed light onto the cusp-core controversy.
In this work, we perform $N$-body simulations of the Sextans dSph galaxy. We 
model its stellar components (the main stellar component $+$ stellar clump) and
the DM halo. We explore different profiles for the DM halo and different
sizes for the starting stellar clump.
We also study the survivability of stellar
substructures in MOdified Newtonian Dynamics (MOND).

This article is organized as follows.
In  Section~\ref{sec:Sextans} we describe some properties of Sextans and their stellar clumps.
The initial conditions for the $N$-body simulations are given
in Section~\ref{sec:Nbody}. The evolution of the substructures is described in Section~\ref{sec:results}. 
Finally, we discuss the implications and give our conclusions in Section~\ref{sec:conclusions}.

\section{Sextans and its kinematic substructures}
\label{sec:Sextans}

Sextans is a dSph galaxy satellite of the Milky Way. It is located at a Galoctocentric
distance of $R_{GC}=86$~kpc \citep{mateo98} and it has a luminosity 
of $L_{V}=(4.37 \pm 1.69)\times10^{5}$~L$_{\odot}$ \citep{lokas09}. It has 
a core radius of $R_{\rm core}=16.6$~arcmin ($\sim0.4$~kpc) and a tidal radius 
$R_{\rm tidal}=160$~arcmin ($\sim4$~kpc) \citep{irwin95}.
Since the majority of the stars in Sextans are older than $10$~Gyr \citep{lee09}, 
\cite{karlsson12} estimated the stellar mass of Sextans to be 
$8.9\pm4.1\times10^5$~M$_{\odot}$ by integrating the light from a 
single $12$~Gyr old stellar population. This mass is consistent with 
the value ($8.5\times10^5$~M$_{\odot}$) obtained by \cite{woo08}. The
corresponding B-band stellar mass-to-light ratio is $\Upsilon_{\star} \approx 2$.

The dynamical mass of Sextans has been estimated from the observed
velocity dispersion profile. Assuming an isotropic velocity dispersion tensor and
a Plummer model for the light distribution in Sextans, 
\cite{kleyna04} inferred that Sextans' total mass within $1$~kpc lies between $3\times10^{7}$
and $1.5\times10^{8}$~M$_{\odot}$. 
Adopting a NFW profile for the dark halo, \cite{walker07} obtained a dynamical mass
of $2.5\times 10^{7}$M$_{\odot}$ within a radius of $600$ pc, whereas 
\cite{strigari07} estimated a total mass of $0.9\times 10^{7}$M$_{\odot}$, also
within the central $600$ pc.
More recently, \cite{lokas09} reported a total mass of
$(4.2 \pm 0.6) \times 10^7$~M$_{\odot}$ assuming an NFW density profile, which implies
a $M/L$ value of $\sim 100$~(M/L)$_{\odot}$. 
All these dynamical studies show that
Sextans is a DM dominated dSph galaxy.

Here, we are interested in the existence of
putative stellar substructures in Sextans.
\cite{kleyna04} reported some evidence for a kinematically and photometrically distinct
population at the Sextans center.
These authors found that the dispersion at the center of Sextans was close to zero, and that such a change
in the dispersion profile coincides with a change in the ratio of 
red horizontal branch stars to blue horizontal
branch stars, i.e. in the stellar populations.
They suggested that this is caused by the sinking 
and gradual dissolution of a stellar cluster at the center of Sextans. 

In a later work, \cite{walker06} presented radial velocities of $294$ possible 
Sextans members. Their larger data set did not confirm \citeauthor{kleyna04}'s (\citeyear{kleyna04}) report of a 
kinematically distinct stellar population at the center of Sextans but 
they did obtain similar evidence 
when they restricted their analysis to a similar (small) number of stars as used 
by \cite{kleyna04}. When considering their full radial velocity sample instead,
\cite{walker06} detected a region near Sextans' 
core radius that is kinematically colder than the overall Sextans sample, with 95\% confidence.
They estimated a substructure luminosity of $3\times 10^{4}$L$_{\odot}$. We will refer to it as
{\it substructure A}.

Recently, \cite{battaglia11} reported nine old stars that share very similar 
spatial location, kinematics and metallicities. The average metallicity of their 9-star 
group is low ([Fe/H]$= -2.6$~dex with a $0.15$ dex scatter), consistent with being
the remnant of a old stellar cluster. This group of stars was taken 
from the six innermost metal-poor stars, which show a cold velocity dispersion of $1.2$~km~s$^{-1}$ 
and an average velocity of $72.5 \pm 1.3$~km~s$^{-1}$. \cite{battaglia11} suggested that 
the number of stars in this substructure (nine stars) is significant with respect 
to the total number of Sextans members for which spectroscopic measurements exist 
($174$ stars). This substructure would account for $5$\% of Sextans' stellar population,
which corresponds to
a luminosity of  $2.2\times 10^{4}$~L$_{\odot}$. We will refer to it as {\it substructure B}.

The present spatial extent of the substructures is very uncertain.
The contours of statistical significance for regions of cold kinematics in Walker et al.
(2006) 
show that substructure A is centered on a location $15$ arcmin north of the Sextans center
and has a radial size of $4$ arcmin ($\sim 100$ pc). On the other hand,
the nine innermost metal-poor stars that constitute the substructure B are found
at $R<0\degree.22$ \citep{battaglia11}, i.e. at 
$\approx 330$~pc from the center of Sextans if we assume a distance to Sextans of $86$~kpc.
In \citeauthor{battaglia11}'s (\citeyear{battaglia11}) data, there are no metal-poor stars at $R<0\degree.1$. This suggests that the substructure B
extends in projected galactocentric radius from $0\degree.1$ to $0\degree.22$ 
(i.e. between $150$ and $330$~pc), indicating
that, in projection,  its center is at $\sim 240$ pc from the
Sextans center, and its radius is $\sim 90$ pc at most.

\section{The N-body Model}
\label{sec:Nbody}
\subsection{Sextans' DM component}
\label{sec:DM}

For our simulations, we constructed a live DM halo with a mass density profile given by
\begin{equation}
\label{eq:halos}
 \rho(r)=\frac{\rho_{0}}{(r/r_{s})^{\gamma} [1+(r/r_{s})^{\alpha}]^{(\beta-\gamma/\alpha)}} \mbox{ ,}
\end{equation}
where $r_{s}$ is the scale radius and $\alpha$, $\beta$ and $\gamma$ define the DM halo's slope. 
This general density profile equation is very useful to define different density profiles. For example, a 
pseudo-isothermal sphere is obtained for $\left( \alpha,\beta,\gamma\right)=(2, 0, 0)$
and an NFW profile is obtained for $\left( \alpha,\beta,\gamma\right)=(1, 3, 1)$.

We explored two different DM radial profiles: a pseudo-isothermal dark halo and an NFW halo. 
We adopted the parameters of \cite{battaglia11} for their best-fitting DM mass modeling of
Sextans, based on their observed line-of-sight velocity dispersion profile. 
For a pseudo-isothermal DM halo, these authors found a core radius $r_s=3$~kpc, 
and a mass of $4\times10^8$~M$_{\odot}$ within the last 
measured point ($\sim2.3$~kpc, assuming a distance to Sextans of $86$~kpc).
In this model, the DM mass within a radius of $0.6$ kpc is $0.9\times 10^{7}$M$_{\odot}$.
For the NFW DM halo, they derived a concentration $c=10$ and a virial mass 
$M_V=2.6\times10^9$~M$_{\odot}$, resulting a DM mass within $0.6$ kpc of
$2.6\times 10^{7}$M$_{\odot}$.

To generate the initial conditions of the DM particles, we used the 
distribution function proposed by \citet{widrow00}, assuming an isotropic velocity dispersion tensor.

\subsection{Sextans' main stellar component}
The main stellar component in Sextans was modeled using the density profile
\begin{equation}
\rho_{*}(r)=\frac{(3-\gamma) M_{*}}{4\pi} \frac{a}{r^{\gamma} (r+a)^{\beta-\gamma}} \mbox{ ,}
\end{equation}
where $M_{*}$ is the total stellar mass and $a$ is the scale radius.
We set 
$M_{\star}\approx 9\times10^5M_{\odot}$,  assuming a typical value of the mass-to-light 
ratio $\Upsilon_{\star}=2$.
We took $\beta = 4$, which corresponds to the
Dehnen models (\citeauthor{dehnen93} \citeyear{dehnen93}; 
\citeauthor{tremaine94} \citeyear{tremaine94}).
In these models, the density declines as $r^{-4}$ at large radii and diverges
in the center as $r^{-\gamma}$. We used $\gamma=3/2$, because it most closely 
resembles the de Vaucouleurs model in surface density.
We took the scale radius for the main stellar component in the Sextans galaxy to 
be $a\approx0.4$~kpc ($16.6$~arcmin, \citeauthor{irwin95} \citeyear{irwin95};
\citeauthor{lokas09} \citeyear{lokas09}). 

\cite{battaglia11} conducted an anisotropy study of 
Sextans and found that the general trend for the best-fitting DM mass model is to have a constant 
anisotropy value close to zero. Thus, the velocity dispersion of the 
main stellar component was taken to be isotropic.

We performed an $N$-body simulation with the DM halo (cored and NFW) and the main stellar component 
together. Each component was found to be stationary for a Hubble time (i.e., the density 
profile of the stellar component and its velocity dispersion stayed approximately constant for 
a Hubble time). 

\subsection{Sextans' stellar clumps}
\label{sec:clump}
Our starting hypothesis is that the cold substructures in Sextans were initially stellar clusters
that are now in the process of {\it very slow} dissolution. The substructures are
the gravitationally unbound remnants of the stellar clusters.
Adopting $\Upsilon_{\star}= 2$, typical for an old stellar population, 
the mass in the substructures lies between  
$M=4.4\times10^{4}$~M$_{\odot}$ for substructure B 
to $\simeq 6 \times10^{4}$~M$_{\odot}$ for substructure A, which are reasonable for
stellar clusters. In all our runs, we assumed, for simplicity, a fixed 
mass of $4.4\times 10^{4}$M$_{\odot}$ for the initial stellar cluster. 

For the initial density profile of the stellar clumps, we used a Plummer model, where
the mass density profile is given by the following equation:
\begin{eqnarray}
\rho_{c}(r) &=& \rho_{0} \left(1+\frac{r^2}{r_c^2}\right)^{-5/2}, \label{rho_plummer}
\end{eqnarray}
\citep{plummer11}. 
We explored different values for the initial core radius $r_{c}$ between $5$ pc,
which corresponds to the size of a typical stellar cluster, and $80$ pc, which is of the order
of the present size of the observed substructures.
Simulations with initial core radius of $r_{c}=5$ pc are aimed to represent a scenario where
the stellar cluster has been caught in the last stage of tidal
disruption. This scenario may present a timing problem because this stage is
expected to proceed on a time-scale of one crossing time of the system and, thus, 
it would be very unlikely to observe them during this phase.
Simulations with an initial radius of $r_{c}=80$ pc correspond to a situation
where the stellar
cluster became unbound immediately after formation due to supernova ejection
of gas (Goodwin 1997).

Without the loss of generality, we set the clumps with an orbit in the $(x,y)$ plane. 
Since we do not know the
orbital parameters of the substructures, we explored different orbits for the clumps
around the Sextans center. We only know lower limits for the semimajor axes of 
the substructures (it is $\gtrsim 400$ pc for substructure A and $\gtrsim 200$ pc
for substructure B). 
Because projection effects lead to an underestimation of the galactocentric distance of the 
objects,
there is a probability of $20\%$ that the substructure B is at a deprojected
distance of $\geq 400$ pc to Sextans' center. Therefore, since either substructure A
or B or both may be on an orbit with a characteristic radius of $400$ pc, we
considered orbits with this size. 
Note that a distance of $\sim 400$~pc corresponds to the core radius
of the main stellar component in Sextans (see Section~\ref{sec:Sextans}).
We also considered the limiting case where the galactocentric distance of the clump is $250$~pc in
order to study a wider range of possible orbits of the clump; this case is relevant for
substructure B.
Since the substructures are not 
necessarily on circular orbits; it is also worthwhile to consider eccentric orbits.

\subsection{The code}
\label{sec:code}
Since the internal two-body relaxation timescales for the three components (clump, 
main stellar component and halo) are much larger than a Hubble time, 
Sextans can be represented as collisionless \citep{binney}. We simulated the 
evolution of the Sextans dwarf galaxy (stellar clump, main stellar component and 
DM halo) using the $N$-body code \scriptsize {SUPERBOX} \normalsize \citep{fellhauer00,bien}. 
\scriptsize {SUPERBOX} \normalsize is a highly efficient particle-mesh, collisionless-dynamics 
code with high resolution sub-grids. 

In our case, \scriptsize {SUPERBOX} \normalsize uses three nested grids centered on the center 
of density of the Sextans dSph galaxy. We used $128^3$ cubic cells for each of the grids. The inner 
grid is meant to resolve the inner region of Sextans and the outer grid (with radii of $100$~kpc 
for all cases) resolves the stars that are stripped away from Sextans' potential. The tidal
field created by the Milky Way was not included. The spatial resolution 
is determined by the number of grid cells per dimension ($N_c$) and the grid radius ($r_{\rm grid}$). 
Then the side length of one grid cell is defined as
$l=\frac{2 r_{\rm grid}}{N_c-4}$. For $N_{c}=128$, the resolution is of the order 
of the typical distance between the particles in the simulation.

\scriptsize {SUPERBOX} \normalsize integrates the equations of motion with a 
leap-frog algorithm, and a constant time step $dt$. We selected a time 
step of $dt=0.1$~Myr in our simulations in order to guarantee that the
energy (for the isolated components) is conserved better than $1\%$.

\section{Results}
\label{sec:results}
Our $N$-body simulations were carried out from an integration time $t=0$ to $t=10$~Gyr.
We used $1\times10^7$ particles to model each of the DM halos, $1\times10^5$ particles
to model the main stellar component, and $1\times10^4$ particles to model the stellar 
clump. The parameters
of our 14 models, from $M1$ to $M14$, are summarized in Table~\ref{table:1}.

\subsection{The cored DM halo case}
\label{sec:res_core}
\begin{figure}
\epsscale{1.}
\plotone{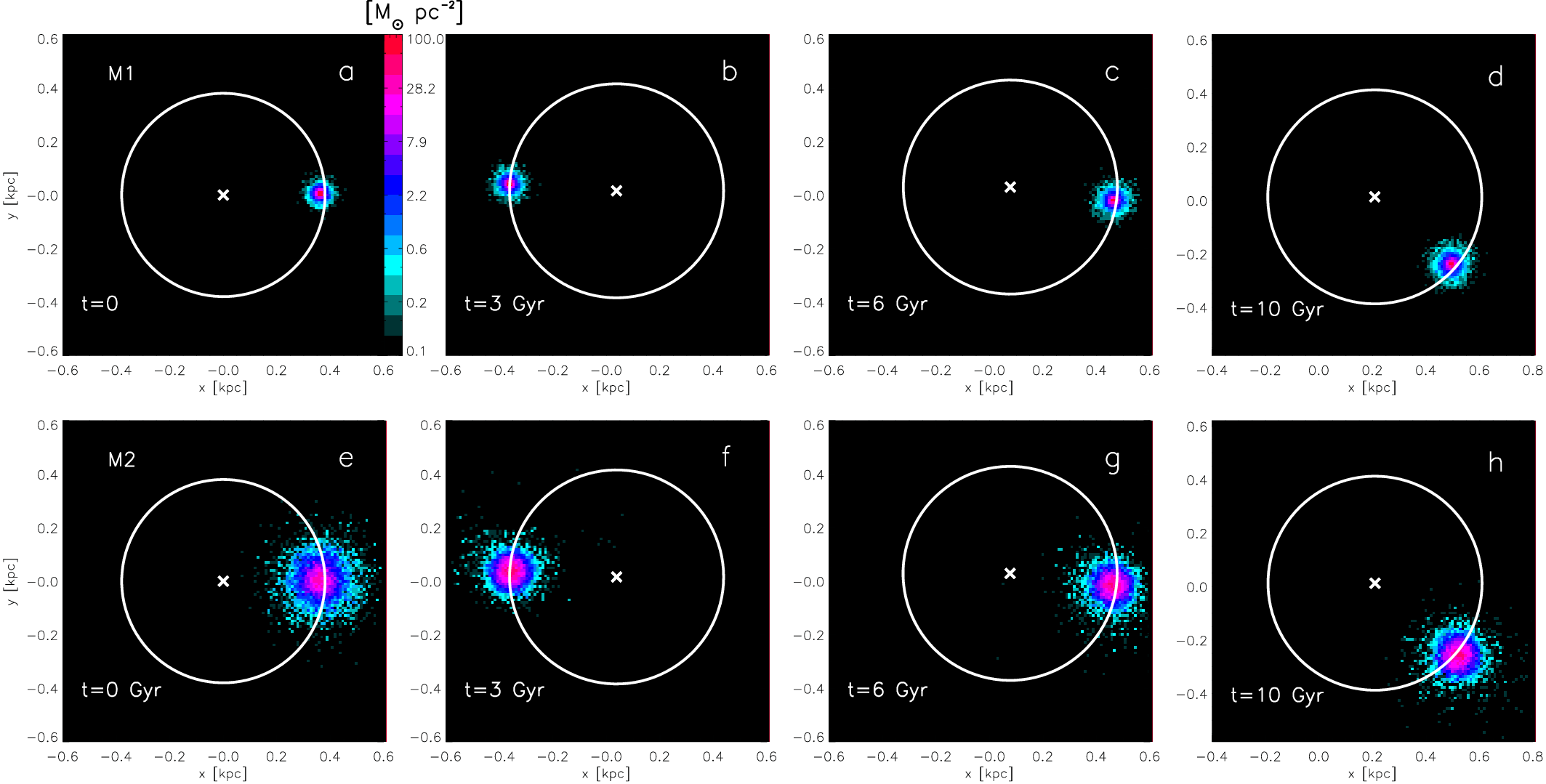}
\caption{Time evolution ($t=0$, $3$, $6$ and $10$ Gyr) of the stellar clump's mass surface density
in the Sextans dSph galaxy. The top panels show the evolution of the clump
in model $M1$ ($r_c=12$ pc), whereas 
the bottom panels show the evolution of the clump in model $M2$ ($r_c=80$ pc). The white 
circle shows the initial orbit with $0.4$~kpc radius. The white cross marks the center 
of Sextans. The main stellar component is not shown.}
\label{fig:FIG1}
\end{figure}

We first consider the evolution of the clump when it is embedded in a pseudo-isothermal DM 
halo, having a core radius $r_s=3$~kpc and a mass within $0.6$~kpc
of $0.9\times10^7$~M$_{\odot}$ (see Section \ref{sec:DM}). 
In order to visualize the evolution of the clump in Sextans, we built a map of 
the surface density (in units of M$_{\odot}$~pc$^{-2}$) of the stellar clump in 
the $(x,y)$-plane.
Figure~\ref{fig:FIG1} shows the temporal evolution of the mass surface density
of the clump in the models $M1$ (clump radius $r_c=12$~pc) and $M2$ (clump radius 
$r_c=80$~pc) for the cored DM halo.
The white circle shows the initial orbit of the stellar clump, and the white cross shows 
the center of Sextans.

Model $M1$ represents a case where the initial clump is very compact; it resembles 
the globular clusters found in the Fornax dSph 
galaxy \citep{mackey03,penarrubia09}, which have core radii ranging from 
$1.4$ to $10$~pc. In $M1$, the stellar cluster remains essentially 
intact over the duration of the run, $10$~Gyr (see Figure~\ref{fig:FIG1}), 
because the tidal heating by the parent galaxy's halo is very low. 

Unbound clumps as extended as the observed substructures (our model $M2$) also remain
essentially unaltered 
for $10$~Gyr (see the bottom panels of Figure~\ref{fig:FIG1}). The physical reason is that 
the underlying potential within the DM core is harmonic and, therefore,
the substructure is long-lived even if it is a gravitationally unbound system
\citep{kleyna03,lora09}. We also
checked that the substructure  can persist for a long time if it is dropped on a circular 
orbit with a radius of $250$ pc (simulation
$M3$). Therefore, we concluded that a dark halo with core radii
of $\gtrsim 3$ kpc may host unbound cold substructures as those observed in Sextans.

\subsection{The NFW DM halo case}
\label{sec:res_nfw}
\begin{figure}
\epsscale{1.}
\plotone{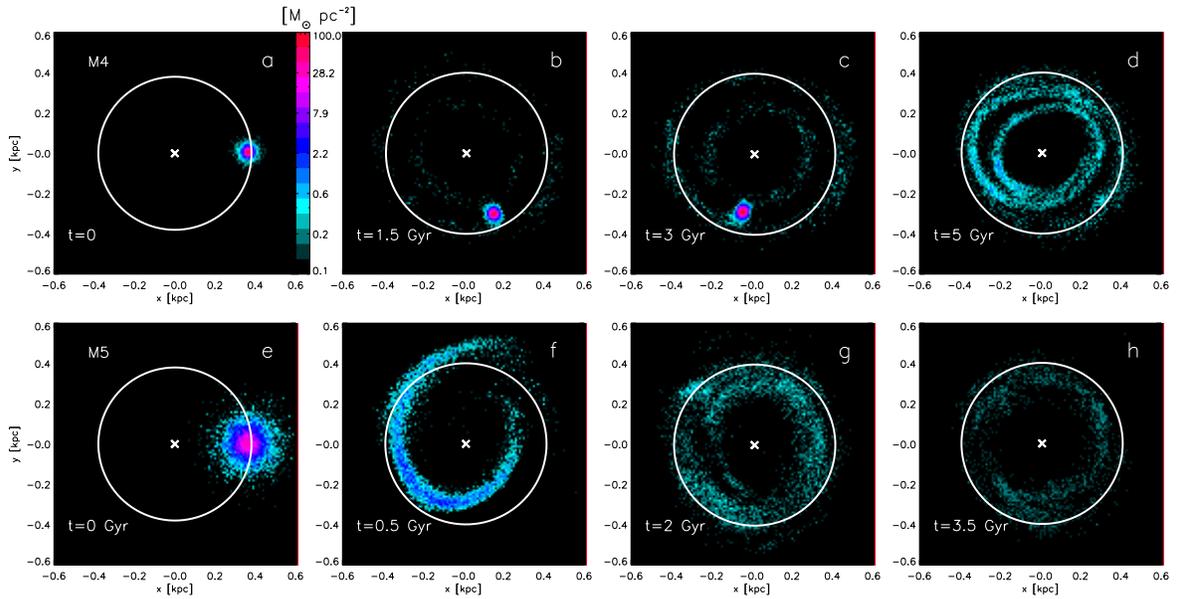}
\caption{The top panels show the mass surface density of Sextans' stellar clump at four different
times ($t=0$, $1.5$, $3$ and $5$ Gyr) in model $M4$ ($r_c=12$ pc). In the bottom panels, we 
show the mass surface density (at $t=0$, $0.5$, $2$ and $3.5$ Gyr) of model $M5$ ($r_c=80$ pc). 
The white circle shows the initial orbit. The white cross marks the center 
of Sextans.}
\label{fig:FIG2}
\end{figure}

The top panels of 
Figure~\ref{fig:FIG2} show the evolution of the mass surface density of the 
stellar clump with initial radius $r_c=12$~pc  (model $M4$), in a NFW halo with a concentration $c=10$ 
and a virial mass of
$M_V=2.6\times10^9$~M$_{\odot}$ (see Section \ref{sec:DM}). For the first $3$~Gyr, the 
stellar clump appears almost unperturbed (see panel $(c)$ of Figure~\ref{fig:FIG2}), 
but after $5$~Gyr the orbital phase mixing dissolves it completely and so only a tidal debris
can be seen in panel $(d)$ of  Figure~\ref{fig:FIG2}. 

The lower panels of Figure~\ref{fig:FIG2} show four snapshots of the substructure when
its initial radius is $r_{c}=80$ pc  (model $M5$). The clump is disrupted in $\sim 0.5$ Gyr 
because of the tidal forces by the parent DM halo. Since this time is much shorter than
its age ($\sim 10$ Gyr), the substructure is short lived. 

\begin{figure}
\epsscale{.8}
\plotone{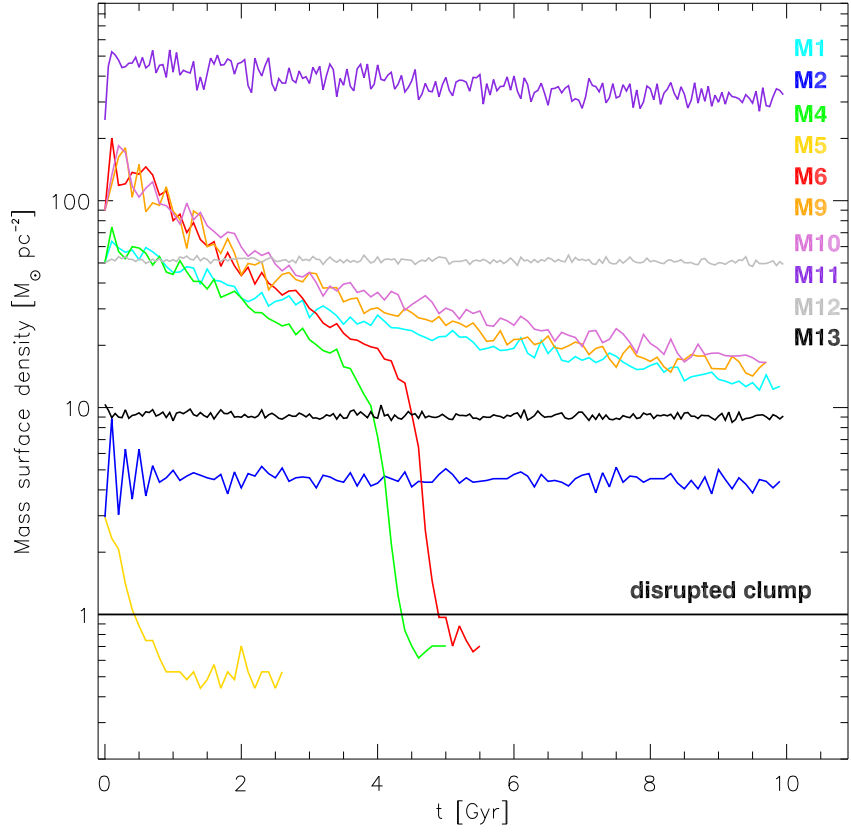}
\caption{Mass surface density of Sextans' stellar clump mapped in the $(x, y)$-plane as
a function of time, for the models quoted at the right margin of the Figure (see Table~\ref{table:1}).}
\label{fig:FIG3}
\end{figure}

At any given time $t$ in the simulation we sample the two-dimensional map searching 
for the $10\times10$~pc size parcel that contains the highest mass (number of clump 
particles).  We define the destruction time as the time at which the parcel with the highest  
mass surface density has reached a value of $\sim1$~M$_{\odot}$~pc$^{-2}$.
When such value is reached, the column density of the clump is so low that 
it would be indistinguishable from Sextans' main stellar component, and would thus be 
undetectable. 
In Figure~\ref{fig:FIG3}, we plot the surface density of this parcel as a function 
of time. We see that the destruction time is $\sim 4.4$~Gyr in model $M4$ and
$\sim 0.45$~Gyr in model $M5$.

\begin{figure}
\epsscale{1.}
\plotone{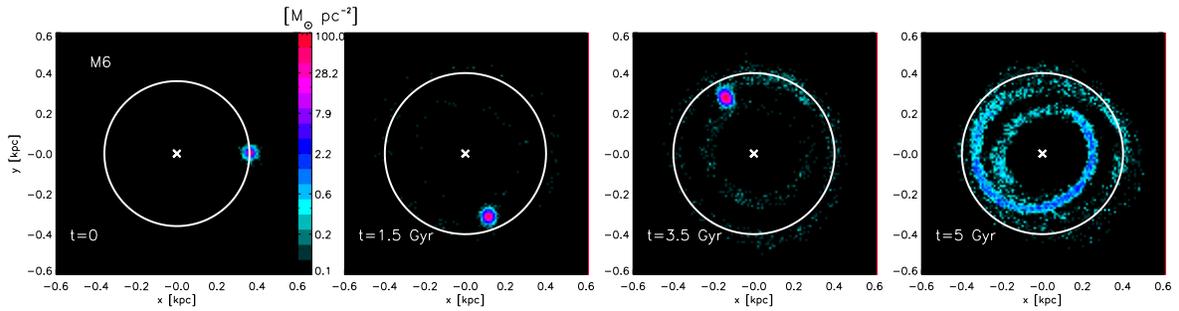}
\caption{Mass surface density of Sextans’ stellar clump in the $(x, y)$-plane in model $M6$ for 
the integration times $t=0$, $1.5$, $3.5$ and $5$~Gyr. The white circle shows the 
initial orbit with $0.4$~kpc radius. The white cross marks the
center of Sextans.}
\label{fig:FIG4}
\end{figure}

In model $M6$, we reduced the initial size of the clump to a radius
$r_c=5$~pc (Figure~\ref{fig:FIG4}). We found that not even with such a
small clump is able to survive for more than $\sim5$~Gyr.

In order to study the effect of the orbital eccentricity on the destruction time,
we ran two simulations ($M7$-$M8$, see Table~\ref{table:1}) where
clump's orbit is non-circular. In 
model $M7$, we set the $5$~pc radius clump in a pure radial orbit with apocenter at $400$~pc 
(see Figure~\ref{fig:FIG6}). 
We found that the clump loses its identity in $4$ Gyr (Figure~\ref{fig:FIG6}). 
In model $M8$, we set the 
$5$~pc radius cluster at a radial distance of $400$ pc
with a tangential velocity twice the circular velocity at that distance 
($v_{y}=2\times v_{c}$). Thus, the orbit is eccentric and its pericenter
is located at $400$ pc.
In this case, the clump is disrupted within $\sim5$~Gyr.
Therefore, it is difficult to explain the existence of cold substructures in 
Sextans at projected distances of $\sim 400$ pc, as substructure A, if an NFW 
profile is adopted for its DM halo. 

\begin{figure}
\epsscale{.8}
\plotone{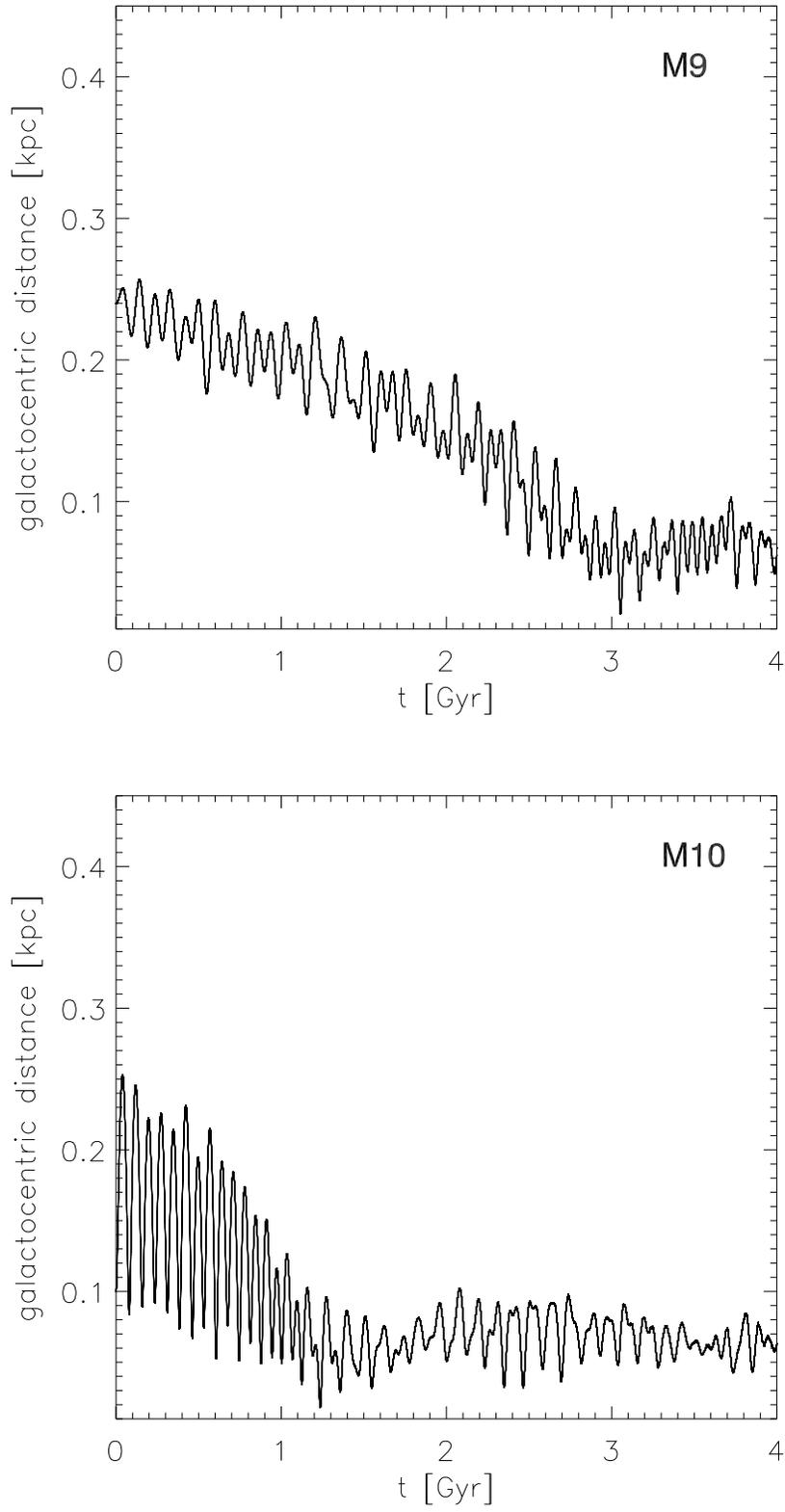}
\caption{Galactocentric distance of the clump as a function of time, for models $M9$ (top
panel) and $M10$ (bottom panel).}
\label{fig:FIG5}
\end{figure}

\begin{figure}
\epsscale{1.}
\plotone{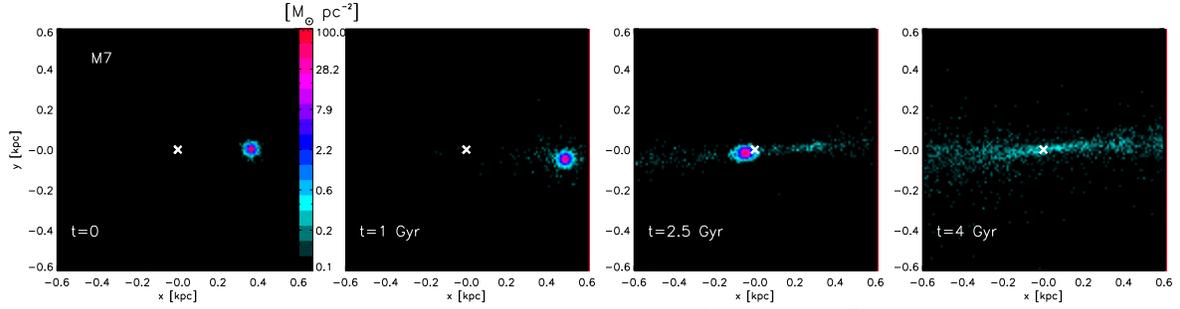}
\caption{Mass surface density of Sextans’ stellar clump in the plane of the orbit, in model $M7$ for 
the integration times $t=0$, $1$, $2.5$ and $4$~Gyr. The white cross marks the center of Sextans.}
\label{fig:FIG6}
\end{figure}

In the case of substructure B, its projected distance is uncertain and we cannot
rule out orbital radii of $\sim 200$ pc or less. In order to explore if this
clump could survive for a significant time,
we ran a simulation ($M9$) where a $5$~pc radius cluster is set on a circular 
orbit with a radius of $250$~pc. 
In this case, the clump starts to lose particles while it spirals to
the center of Sextans DM potential due to dynamical friction. 
At $t\approx3$~Gyr, the stellar clump reaches
the center of Sextans, and it keeps orbiting very close to 
the center until the end of the simulation (see Figure~\ref{fig:FIG5}). 
As a consequence, the stellar clump survives as a central star cluster. 
It has to be noticed that the dissolution of the clump in this case is 
similar to  model~$M6$ only during the first $\sim3$~Gyr.

The evolution of the clump occurs in a similar
manner in model $M10$, where the orbit of the $5$~pc radius clump 
has its apocenter at $250$~pc and its pericenter at $100$~pc.
The clump spirals to the center of Sextans and loses mass, until the density of the 
clump decreases by a factor of $\sim 2$ at $t\approx1$~Gyr (see Figure~\ref{fig:FIG5}).
After this time, the orbital radius of the clump lies between $30$ and $90$ pc
and its dissolution proceeds in a very slowly way.

We conclude that cold substructures having initial orbital radii of $\sim 400$ pc
would not survive in an NFW halo. Only substructures with initial orbital radii $\lesssim 300$ 
pc could survive but they should be located close to the Sextans center (distances $<100$ pc)
because of the orbital decay due to dynamical friction. Since substructure A is at a projected
distance of $400$ pc, it is difficult to understand how it survived against
mixing in an NFW halo. A more accurate determination of the projected distance of
substructure B would be very important to constrain the models further.

%
\subsection{The case of MOND} 
\label{sec:mond}
It is interesting to explore if the gravitational potential predicted in
MOND could explain the survival of cold substructures. 
To do so, we 
followed a similar treatment as in \cite{sanchez-lora} for the UMi dSph galaxy.
In the MOND framework, the gravitational potential that describes the force acting on a 
star in Sextans follows the modified Poisson equation of \cite{bekestein}
\begin{equation}
 \nabla \cdot [\mu (x) \nabla \Phi]=4 \pi G \rho \mbox{,}
\label{eq:mond}
\end{equation}
where $x=|\nabla \Phi|/a_{0}$, $a_0\simeq1.2\times10^{-8}$~cm~s$^{-2}$ is the universal
acceleration constant of the MOND theory, and $\mu (x)$ is the interpolating function, 
which runs smoothly from $\mu (x)=x$ at $x\ll1$ to $\mu (x)=1$ at $x\gg1$. The
differential equation 
(\ref{eq:mond}) must be solved with the boundary condition 
$\nabla \phi \rightarrow -\textbf{g}_{E}$, where $\textbf{g}_{E}$ is the external gravity 
acting on Sextans and has a magnitude ${g}_{E}=V^2/R_{GC}$. $V$ is the Galactic rotational 
velocity at $R_{GC}$ which coincides with the asymptotic rotation velocity $V_{\infty}$ 
for the Milky Way. We set this value to $V_{\infty}=170$~km~s$^{-1}$ which is obtained by 
adopting a mass model for the Milky Way under MOND \citep{famaey05,sanchez}.

A star in the stellar clump of Sextans feels the external acceleration created by the Milky 
Way (denoted by $\textbf{g}_{E}$), the acceleration generated by Sextans smooth stellar component 
($\textbf{g}_{I}$), and the acceleration generated by all other stars that form the stellar 
clump ($\textbf{g}_{int}$), and thus all must be taken in consideration.

Since the circular velocity of a test particle at the stellar core radius 
($r_{*}\simeq0.4$~kpc) of Sextans is $\sim5.9$~km s$^{-1}$, then the characteristic internal 
acceleration $[v_c(r_*)]^2/r_*\simeq2.9\times10^{-10}$~cm s$^{-2}$. This is much smaller than 
MOND's characteristic acceleration $a_{0}\simeq1.2\times10^{-8}$~cm s$^{-2}$. Moreover, the 
external acceleration, $V^2_{\infty}/R_{GC}\simeq0.11\times10^{-8}$~cm s$^{-2}$ is also much 
smaller than $a_{0}$. We can conclude then that the Sextans internal dynamics lies 
deep in the MOND regime. The ratio between the internal acceleration at Sextans' stellar 
core radius and the external acceleration (${g}_{I}/g_{E}\approx0.26$) tells us that the 
dynamics in Sextans is dominated by the external field ($g_{E}\gg g_{I}$). 

\cite{sanchez} studied the dSph galaxies of the Milky Way under MOND 
and compared the results with DM halos. For Sextans, they obtained a high value of (M/L) of 
$\sim7.5-36$ assuming that the external field is dominant in this galaxy. 
\cite{angus08} analyzed the line-of-sight velocity dispersion as a function of radius for 
eight Milky Way dSph galaxies and calculated the mass-to-light ratio 
in the MONDian regime through a Jeans analysis. He found that Sextans 
requires a rather high mass-to-light ratio of $9.2$. We adopted this value here.

\begin{figure}
\epsscale{1.}
\plotone{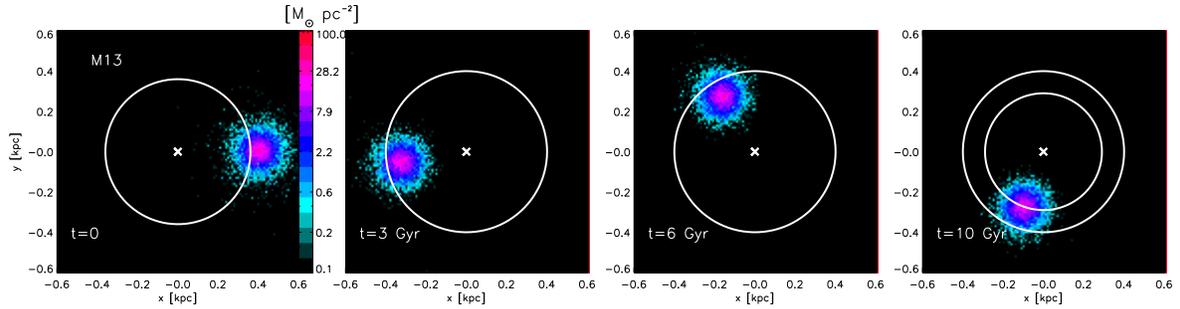}
\caption{Mass surface density of Sextans’ stellar clump in the $(x, y)$-plane in model $M13$ for 
the integration times $t=0$, $3$, $6$ and $10$~Gyr.
The outer white circle shows the initial orbit of the stellar clump, with $0.4$~kpc radius. The inner circle shows
the final orbit with a radius of $0.29$~kpc.}
\label{fig:FIG7}
\end{figure}

We carried out $N$-body simulations under the MOND approximation with the code 
described in Section \ref{sec:Nbody} starting with clumps of different radii ($r_{c}=12$,
$35$ and $80$ pc, which correspond to models $M11$, $M12$ and $M13$, respectively). 
The parameters of the different models are summarized in Tables \ref{table:1}
and~\ref{table:1B}.
The clump survives for more than $10$ Gyr in the three cases 
(see Figure~\ref{fig:FIG3}).
It is interesting to note that when the initial stellar clump is extended ($r_{c}=80$ pc), 
it spirals to the center of  Sextans; at $t=10$~Gyr it orbits within $\sim 0.3$~kpc.
This is clearly seen in Figure~\ref{fig:FIG7}.

Finally, we ran a simulation ($M14$) of the $80$~pc radius clump in a circular orbit with 
a galactocentric distance of $250$~pc. Also in this case the clump remains undisturbed for 
$10$~Gyr. In this case, the clump also spirals to the center of Sextans, and at $t=10$~Gyr 
it orbits within $\sim 0.2$~kpc.


\section{Concluding remarks} 
\label{sec:conclusions}

Using $N$-body simulations, we studied the survival
of cold kinematic substructures in the DM halo of the Sextans dwarf galaxy against
phase mixing.
We compared the evolution of  substructures when the dark halo has a core and when
the dark halo follows the NFW profile, those having
the parameters derived by \cite{battaglia11} to explain the projected velocity dispersion
profile. 
We found that the core in the pseudoisothermal model is large enough to make the potential
almost harmonic and, thus, to
guarantee the survival of substructures. Even if the clump is initially very extended ($r_c=80$~pc), 
it easily survives for $10$~Gyr.
We conclude that 
the stellar clump in Sextans is in agreement with a cored DM halo.

On the contrary, stellar clumps orbiting with semi-major axes of $\sim 400$ pc
and initial Plummer radii between $12$ and $80$~pc
are destroyed if they are embedded in the NFW DM halo.
Not even a stellar clump with a small Plummer radius of $r_c=5$~pc (model $M6$) 
can survive in such a NFW DM halo. 
Stellar clumps initially orbiting at a radius $\leq 250$ pc from the Sextans center
spiral to the center due to dynamical friction and, as a consequence,
phase mixing is reduced. Clumps in these orbits may survive but would merge
forming a central star cluster at the center of Sextans.

It has to be noted that we cannot rule out a scenario where
the DM profile was initially cuspy and evolved to a cored halo.
For instance, energy feedback from supernova explosions and stellar winds, 
may lead severe gravitational potential fluctuations, which may reduce the 
central mass density of dwarf galaxies (e.g., \cite{mashchenko08}).  Similarly, \cite{pasetto10} found that initial 
cuspy DM profiles flatten with time as a result of star formation, which would explain the 
observations without contradicting a cuspy DM profile. On the other hand, \cite{penarrubia12} 
show some difficulties with the fine-tuning of the scenarios just mentioned. \cite{goerdt10} 
discuss that the transfer of energy from sinking massive objects may destroy central cusps.

As a last point, we investigated whether or not the clump in Sextans could survive in 
the MONDian framework. We found that even a stellar clump with $r_c=80$~pc remains undisturbed 
for a Hubble time and slowly spirals to the center of Sextans. 
However, it has to be noticed that the adopted MOND value for the stellar
mass-to-light ratio of $M/L_{V}=9.2$, which was derived by
\cite{angus08} to explain the observed velocity dispersion profile of Sextans, 
is very high and inconsistent with the properties expected from an old purely stellar 
population.


\section*{Acknowledgments}
We wish to thank the anonymous referee for very useful comments that 
greatly improved the content of this paper.
V.L. gratefully acknowledges support from an Alexander von Humboldt Foundation fellowship and
the FRONTIER grant. F.J.S.S. acknowledges financial support from CONACyT
project 165584 and PAPIIT project IN106212.


\clearpage

\begin{table}
\centering
\caption{Parameters of the models}
\medskip
\begin{tabular}{@{}ccccc@{}}
\hline
Model & Halo & $r_{c}$ & Surviving time & Type of orbit \\
& profile$^{a}$ & [pc] & [Gyr]&\\
\hline
\hline
& & & &\\
M1 & ISO & $12$ & $>10$ & circular orbit of radius of $400$ pc \\

M2 & ISO & $80$ & $>10$ & circular orbit of radius of $400$ pc \\

M3 & ISO & $80$ & $>10$ & circular orbit of radius of $250$ pc \\

M4 & NFW & $12$ & $\sim4.4$ & circular orbit of radius of $400$ pc \\

M5 & NFW& $80$ & $\sim0.45$ & circular orbit of radius of $400$ pc \\

M6 & NFW & $5$ & $\sim 5$ & circular orbit of radius of $400$ pc \\

M7 & NFW & $5$ & $\sim 4$ & radial, apocenter at $400$ pc \\

M8 & NFW & $5$ & $\sim 5$ & eccentric, pericenter at $400$ pc, \\

& & & & velocity at pericenter: $v_{x}=0$, $v_{y}=2v_{c}$ \\

M9 & NFW & $5$ & $>10^{b}$& circular orbit of radius of $250$ pc \\

M10 & NFW & $5$ & $>10^{c}$ & eccentric, apocenter at $250$ pc\\
& & & & and pericenter at $100$ pc \\
M11 & MOND & $12$ & $>10$ & circular orbit of radius of $400$ pc \\

M12 & MOND & $35$ & $>10$ & circular orbit of radius of $400$ pc \\

M13 & MOND & $80$ & $>10$ & circular orbit of radius of $400$ pc \\

M14 & MOND & $80$ & $>10$ & circular orbit of radius of $250$ pc \\
\hline
\end{tabular}
\vskip .3cm
$^{a}$ISO refers to the pseudo-isothermal profile.\\
$^{b}$The orbit decays to the Sextans center in $\sim 3$ Gyr.\\
$^{c}$The orbit decays to the Sextans center in $\sim 1$ Gyr.
\label{table:1}
\end{table}


\begin{table}
 \centering
 \caption{Parameters of the Sextans dSph and its stellar clump}
 \medskip
 \begin{tabular}{@{}ccccccccc@{}}
 \hline
Sextans &  D     & $L_{V}$         & r$_{*}$  & $M_{*}$             &$M(<r_{*})$   & $v_c(r_{*})$ & $g_{I}/g_{E}$\\
        &  [kpc] & [L$_{\odot}$]   & [arcmin] &[M$_{\odot}$]    & [M$_{\odot}$]    &  [km s$^{-1}$]&  at $r_{*}$ \\
        & $86$   & $4.37\times10^5$& $16.6$   & $8.7\times10^5$ & $3.18\times10^5$ &    $5.9$      & $0.26$     \\
 &  &    &  &  &  & \\

Clump  &  Semi-major   & $L_{V}$         & r$_{h}$ & $M$            &             & $v_c(r_{h})$   &  $g_{int}/g_{E}$ \\
       & axis  [kpc]  & [L$_{\odot}$]   & [pc]    &[M$_{\odot}$]   &             &  [km s$^{-1}$] &    at $r_{h}$    \\
Small clump      & $0.4$       & $2.2\times10^4$ & $15.6$  & $2.02\times10^5$&             &     $1.3$      &     $0.35$  \\
 &  &    &  &  &  & \\

Big clump      & $0.4$       & $2.2\times10^4$ & $45.5$  & $2.02\times10^5$&             &     $1.8$      &     $0.10$  \\

  \hline
  \end{tabular}
  \label{table:1B}
 \end{table}

\end{document}